\documentclass[%
 reprint,
superscriptaddress,
 amsmath,amssymb,
 aps,
]{revtex4-2}

\usepackage{svg}
\usepackage{graphicx}% Include figure files
\usepackage{dcolumn}% Align table columns on decimal point
\usepackage{bm}% bold math
\usepackage{siunitx}
\usepackage{amsmath}
\usepackage{mathtools}
\usepackage{braket}
\usepackage{appendix}
\DeclarePairedDelimiter\floor{\lfloor}{\rfloor}

\newcommand{\dx}[2]{\frac{\mathrm{d}#1}{\mathrm{d}#2} }

\newcommand{\tr}[1]{\mathrm{tr\left[#1\right]}}
\newcommand{\ptr}[2]{\mathrm{tr_{#2}\left(#1\right)}}

\begin{document}

\preprint{APS}%/123-QED}

\title{Coherent laser cooling with trains of ultrashort laser pulses}% Force line breaks with \\
% \thanks{A footnote to the article title}%

\author{J. Malamant}
\affiliation{Department of Physics, University of Oslo, Sem Saelandsvei 24, 0371 Oslo, Norway}

\author{N. Gusakova}
\affiliation{Department of Physics, NTNU, Norwegian University of Science and Technology, Trondheim, Norway}
\affiliation{Physics Department, CERN, 1211~Geneva~23, Switzerland}

\author{H. Sandaker}
\affiliation{Department of Physics, University of Oslo, Sem Saelandsvei 24, 0371 Oslo, Norway}

\author{I.~T.~Sorokina}
\affiliation{Department of Physics, NTNU, Norwegian University of Science and Technology, Trondheim, Norway}

\author{D. Comparat}
\affiliation{Universit\'e Paris-Saclay, CNRS, Laboratoire Aim\'e Cotton, 91405, Orsay, France.}

\author{A. Camper}
\email[]{Corresponding author: antoine.camper@fys.uio.no}
\affiliation{Department of Physics, University of Oslo, Sem Saelandsvei 24, 0371 Oslo, Norway}

 %Lines break automatically or can be forced with \\
% \author{Second Author}%
%  \email{Second.Author@institution.edu}
% \affiliation{%
%  Authors' institution and/or address\\
%  This line break forced with \textbackslash\textbackslash
% }%

% % \collaboration{MUSO Collaboration}%\noaffiliation

% \author{Charlie Author}
%  \homepage{http://www.Second.institution.edu/~Charlie.Author}
% \affiliation{
%  Second institution and/or address\\
%  This line break forced% with \\
% }%
% \affiliation{
%  Third institution, the second for Charlie Author
% }%
% \author{Delta Author}
% \affiliation{%
%  Authors' institution and/or address\\
%  This line break forced with \textbackslash\textbackslash
% }%

% \collaboration{CLEO Collaboration}%\noaffiliation

\date{\today}% It is always \today, today,
             %  but any date may be explicitly specified

\begin{abstract}
We propose to extend coherent laser cooling from narrow-band to broad-band transitions by using trains of ultrashort broadband pulses. We study analytically two possible methods to reduce the momentum spread of a distribution by several units of photon momentum in a single spontaneous emission lifetime. We report on numerical simulations of one-dimensional laser cooling of a two-level system with realistic parameters. The technique introduced here is of high interest for efficient laser cooling of fast species with short lifetime such as positronium.
\end{abstract}

%\keywords{Suggested keywords}%Use showkeys class option if keyword
                              %display desired
\maketitle

\section{Introduction}

Laser cooling is a well-established technique to increase the phase-space density of an ensemble of particles \cite{Ketterle:92,Chalony:11}. It is based on the interplay of two processes. Internal state laser excitation is used to reduce the particles' momentum spread without changing entropy. Then, spontaneous emission dissipates entropy such that the number of particles with low momentum and in the same internal state increases. For a momentum distribution of N internal level particles which are initially all in the ground state, the maximum of the position--momentum phase--space density can at most be increased by a factor N without dissipating entropy by populating all N levels within the same elementary part of the position--momentum space \cite{Chaneliere18}. The time it takes to compress the momentum spread depends on the ability of the laser to move the particles with high absolute momentum from the ground state to one of the (N-1) available excited states with low absolute momentum, in a time short in front of the shortest spontaneous emission lifetime of all excited states. 

For species prone to annihilation (such as positronium~\cite{Gloggler24,shu2023laser} the bound state of an electron and a positron), radioactive decay, photo-ionization or molecular dissociation, interaction with the cooling laser is limited in time by the fast disintegration of the system. For those that are also usually produced in rather small amounts compared to stable elements, a technique allowing to cool the whole momentum distribution is highly desirable. Recently, the development of coherent laser cooling techniques \cite{Corder15,Bartolotta18} has demonstrated that it is possible to hasten the process of laser cooling. In a similar way to stimulated focusing and deflection of atomic beams \cite{Goepfert97}, these techniques make use of a succession of absorption and stimulated emission to compress the momentum spread in an accelerated way compared to standard Doppler laser cooling. Coherent laser cooling has been designed for narrow line transitions and is currently restricted to narrow momentum distributions by the use of modulated narrow-band lasers. Here, we propose to extend coherent laser cooling to broad line transitions and large Doppler profiles by using trains of ultrashort broadband pulses to drive the process. We discuss two different methods to manipulate the momentum distribution and reduce its rms (root-mean-square) in the one-dimensional case. The first approach consists in moving the positive and negative momentum halves of the Doppler profile towards zero momentum. The second approach is designed to transform the initial Maxwell-Boltzmann distribution into another Maxwell-Boltzmann distribution with half the initial momentum rms. For both methods, we derive the analytical formula predicting the evolution of the momentum rms as function of the number of pulses in the train in the case of ideal population transfer. We further test the robustness of one of the approaches by performing numerical simulations for a series of trains of pulses interacting with a Maxwell-Boltzmann distribution using parameters achievable with today's laser technology and including various realistic effects such as non-ideal population transfer, spontaneous emission and photo-ionization from the excited states.

\begin{figure*}
	\centering
    \includegraphics[width=\linewidth]{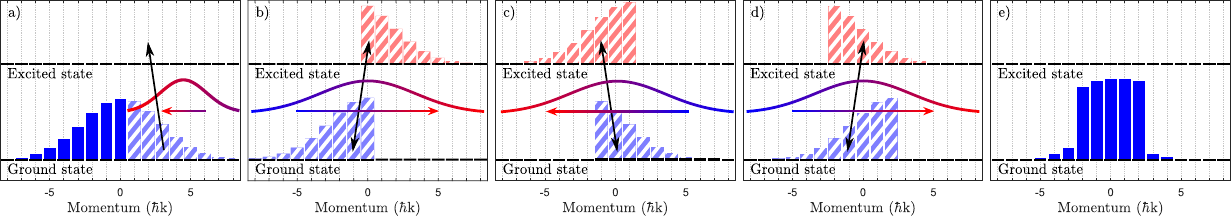}
    \caption{\label{fig:principle1} Illustration of coherent laser cooling with \textit{Identical Pulses} on a distribution with an initial momentum rms of $\simeq\;$5\;$\hbar k$. Colors reflect the electronic internal level. Blue (respectively red) corresponds to the ground (resp. excited) state. a) Initial distribution. From b) to d): distributions corresponding to $M = 1$ to $M = 3$ in equation~\ref{eq:analytical2}. e) Distribution after relaxation via spontaneous emission. The color gradient in the Gaussian shapes represent the laser frequency chirp of pulses that could be used for this type of adiabatic population transfer. The arrow under the Gaussian shapes represent the direction of propagation of the laser pulses. The black arrows indicate the direction of population transfer.}
\end{figure*}

\section{\textit{Identical Pulses} method}
In this section, we first consider an ideal case based on a N$ = 2$ level system with an arbitrary one--dimensional distribution $\rho_{0}$ describing the initial momentum population of atoms in the ground state. First, we consider the \textit{Identical Pulses} method which, after an initial desymmetrization step, consists in swapping the populations in the ground and excited states across the entire momentum distribution to compress the momentum rms of the distribution as much as possible (see Fig.~\ref{fig:principle1}). The population transport can be realized with a series of identical pulses after desymmetrization. This is the main strength of this method and we therefore refer to it as \textit{Identical Pulses}. Typically, $\rho_{0}(n)$ is a Maxwell-Boltzmann distribution sampled on integer multiples of $\hbar k$ (one unit of photon momentum) but the derivations reported in this manuscript are general and apply to any type of distribution. 

The system is initially in the ground state and we consider ideal population transfer for all pulses in the train. As a first step, we assume that the length of the train is negligible in front of $\tau_{SE}$ (the spontaneous emission lifetime of the excited state) and therefore assume no transfer of population by spontaneous emission. In the train, the first pulse (or desymmetrization pulse) is particular and is shaped to transfer only half of the momentum distribution (see Fig.\;\ref{fig:principle1}~a) from the ground to the excited state and remove 1$\;\hbar k$ to all particles with strictly positive momentum ($n > 0$) by absorption. The precise shaping of such a pulse is outside the scope of the present work but can be approached by optimization using for example the Krotov algorithm~\cite{Krotov-2019}. 
%In principle, the desymmetrization pulse can be skipped under the reasonable assumption that the momentum distribution is spatially distributed so that the classes of momentum with highest absolute value can be excited by the laser before the classes of momentum with lowest momentum. In this case, p
Pairs of counter-propagating pulses are synchronized to reach the center of the distribution simultaneously, in a fashion reminiscent of SWAP cooling \cite{Bartolotta18}. The second pulse in the train propagates in the direction opposite to the first one and interacts with the whole momentum distribution. The particles with negative or null momentum are transferred from the ground to the excited state and gain 1$\;\hbar k$ by absorption while the particles that were in the excited state lose 1$\;\hbar k$ by stimulated emission and end up in the ground state. A train of pulses with alternating direction of propagation is then used to interact with the whole distribution and repeat the population transfer towards the lower class of velocities. 
%The optimal number of pulses in this train minimizes the momentum spread calculated by summing over the internal states.
Next, we derive the analytical formula predicting the evolution of the momentum spread as a function of the number of pulses for the scheme we have just described.

To describe the most general case, we introduce $\rho_{1}(n)$, the initial distribution of momentum for atoms in the excited state, so $\sum_{n}(\rho_{0}(n)+\rho_{1}(n)) = 1$. Being in the ``\textit{wrong}" initial state, the evolution of the ($\rho_{1}(n)$) momentum is opposite to the one of the particles in the ``\textit{right}" initial state ($\rho_{0}(n)$). The evolution of $p_{rms}(M)$ (the momentum rms in $\hbar k$ units) with $M$ (the number of pulses in the train) can be expressed for $M \geq 1$ as:
\begin{equation}\label{eq:analytical2}
\begin{aligned}
    p_{rms}^2(M) =& \sum_{n\leq0}(n+(M-1))^2\rho_{0}(n)+\sum_{n > 0}(n-M)^2\rho_{0}(n)\\
    +& \sum_{n\leq0}(n-(M-1))^2\rho_{1}(n)+\sum_{n > 0}(n+M)^2\rho_{1}(n)
\end{aligned}
\end{equation}
Equation~(\ref{eq:analytical2}) describes a hyperbolic behaviour with a minimum $p_{rms,m}$ reached for M$_{m} = \floor{M_{mc}+\frac{1}{2}}$ pulses. 
\begin{equation}\label{eq:hyperbola2}
    p_{rms,m} = \sqrt{p_{rms,0}^2+M_m\left(M_m-2 M_{mc}\right)+M_{ms}}
\end{equation}
$p_{rms,0}$ (the momentum rms of the initial distribution), $M_{mc}$ and $M_{ms}$ are given in Appendix. For symmetrical momentum distributions, we note that $M_{ms} = M_{mc}$. 
Equation~(\ref{eq:hyperbola2}) is very general and can be used for any discrete momentum distribution to evaluate how fast and by how much the standard deviation of a distribution can be reduced under the ideal conditions assumed to derive it. We note that, in the absence of spontaneous emission, the entropy of the system is unchanged throughout the laser-particle interaction. In case of an initial Maxwell--Boltzmann distribution of atoms all in the ground state, it is possible to show that:
\begin{equation}
    \lim_{p_{rms,0}\to\infty} \frac{p_{rms,0}}{p_{rms,m}} = \frac{1}{\sqrt{1-\frac{2}{\pi}}} \simeq 1.66
\end{equation}
For high enough initial spreads, this technique allows to reach a momentum spread compression factor close to 1.66 with a train made of  $\simeq p_{rms,0}\sqrt{\frac{2}{\pi}}$ pulses.%$\floor{p_{rms,0}\sqrt{\frac{2}{\pi}}+\frac{1}{2}}$ pulses.

\begin{figure*}
	\centering
	\includegraphics[width=1\linewidth]{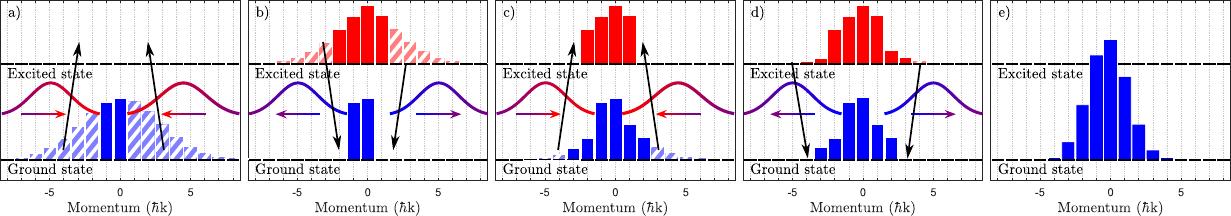}
    \caption{\label{fig:principle2} Illustration of coherent laser cooling with \textit{Optimal Transport} on a distribution with an initial momentum rms of $\simeq\;$5$\;\hbar k$. Colors reflect the electronic internal level. Blue (respectively red) corresponds to the ground (resp. excited) state. a) Initial distribution. Distributions correspond to b) $M = 1$, c) $M = 2$ and d) $M = 3$ in equation\;\ref{eq:notchedanalytical}. e) Distribution after relaxation via spontaneous emission.}
\end{figure*}

The hyperbolic shape of $p_{rms}(M)$ and the existence of an absolute minimum (see Fig.~\ref{fig:sigmaVsN}) is inherent to the specific way the different parts of the distribution are manipulated with the \textit{Identical Pulses}. In particular, the ground state particles with the smallest initial absolute momentum see their momentum increase in absolute value. As a result, the momentum spread compression $\frac{p_{rms,0}}{p_{rms,m}}$ is smaller than the optimal factor 2 allowed for two--level systems \cite{Chaneliere18}. 

\section{\textit{Optimal transport} method}
In order to approach the factor 2 in momentum spread compression predicted by theory for a two--level system, we introduce a second method in which only the outermost parts of the momentum distribution are translated towards its center at each step in order to keep a Gaussian shape for the momentum distribution (see Fig.~\ref{fig:principle2}). This method is inspired by the linear rearrangement solution to the Monge-Kantorovich problem consisting in finding the optimal transport function dividing the rms of a Maxwell-Boltzmann distribution by a factor 2 with a linear cost function. %is reminiscent of solutions found in Optimal Transport problems. 
We therefore refer to it as the \textit{Optimal Transport} method. The general idea is to use pairs of counter propagating pulses which do not interact with classes of momentum where both internal states are populated. In a pair of counter propagating pulses, the two pulses are almost identical. Two consecutive pairs of pulses are however different. The first pair of pulses transfers the atoms with momentum larger than 1$\;\hbar k$ or smaller than -2$\;\hbar k$ from the ground to the excited state with a reduction of absolute momentum by 1$\;\hbar k$. The next pair of pulses transfer the atoms with momentum larger than 2$\;\hbar k$ or smaller than -3$\;\hbar k$ from the excited to the ground states removing again 1$\;\hbar k$ from the absolute momentum of all particles etc. 
In this case, the evolution of the momentum spread with the number of pulses can be expressed for $M \geq 1$ as:
\begin{equation}\label{eq:notchedanalytical}
\begin{aligned}
    \tilde{p}^{2}_{rms}(M) =& \sum_{n = -M+1}^{M-1}n^2\left(\rho_{0}(2n)+\rho_{0}(2n+1)\right)\\
    &+\sum_{n \geq M}n^2(\rho_{0}(n+M)+\rho_{1}(n-M))\\
    &+\sum_{n \leq -M}n^2(\rho_{0}(n-M+1)+\rho_{1}(n+M-1))
\end{aligned}
\end{equation}
% $\sigma_{e}(N)$ is given in $\hbar$k units. 
For $\rho_{1}(n) = 0$, $\tilde{p}_{rms}(M)$ is monotonically decreasing (see Fig.~\ref{fig:sigmaVsN}) and
% In the limit $N\rightarrow +\infty$, 
\begin{equation*}
\lim_{M\to\infty} \tilde{p}_{rms}(M) =  \sqrt{\frac{\tilde{p}_{rms,0}^2}{4}-\sum_{-\infty, n odd}^{\infty}{\frac{2n-1}{4}\rho_0(2n+1)}}
\end{equation*}
For a number of pulses M (see Fig.~\ref{fig:principle2}) large in front of the initial momentum rms, the momentum compression factor is very close to 2. For an initial distribution such that $\forall n \in \mathbf{Z}, \rho_{0}(2n+1) = 0$, the theoretical minimum $\frac{\tilde{p}_{rms,0}}{2}$ is reached. The use of \textit{Optimal Transport} allows to reach higher momentum spread compression factors at the cost of a more complicated population manipulation. The optimal number of pulses in \textit{Optimal Transport} is a compromise between the level of momentum spread compression and the total length of the train that should still be short in front of the relaxation time constant of the system. 

In Fig.~\ref{fig:principle2}.e, the momentum distribution is slightly asymmetric. This is due to the discretization of the momentum space on integers of $\hbar k$. If instead the momentum space is discretized on the half--integer multiples of $\hbar k$, \textit{Optimal Transport} can be composed of pairs of exactly identical pulses addressing classes of momentum larger than 1.5$\;\hbar k$, 2.5$\;\hbar k$ and so on. In this case, the final distribution is perfectly symmetric. 

\section{Multiple trains of short pulses}
\label{sec:multipletrains}
So far, we discussed the evolution of the momentum distribution as function of the number of pulses (one sub-panel in Fig.~\ref{fig:principle1} and ~\ref{fig:principle2}) in a single train (all panels in Fig.~\ref{fig:principle1} and ~\ref{fig:principle2}). We will now discuss the possibility to use several trains of pulses to further reduce the momentum spread of the distribution. After a train with optimal number of pulses has interacted with the ensemble of particles, the system starts to relax to the ground state. Once all atoms are back to the ground state, it is possible to send a second train of pulses and further reduce the momentum spread. The relaxation to ground state through spontaneous emission can be modeled as an exponential decay with a certain probability to gain or lose at most 1$\;\hbar k$. Starting from the $\rho_0(n)$ and $\rho_1(n)$, exponential decay after time $\tau$ is simulated by replacing $\rho_0(n)$ and $\rho_1(n)$ with~\cite{Bloch08}:
\begin{equation*}
\begin{aligned}
    \Bar{\rho}_0(n,\tau) =& \rho_0(n) + 0.2\cdot\rho_1(n-1)\cdot e^{-\tau/\tau_{SE}}\\
    &+ 0.6\cdot\rho_1(n)\cdot e^{-\tau/\tau_{SE}}\\ 
    &+ 0.2\cdot\rho_1(n+1)\cdot e^{-\tau/\tau_{SE}}\\
    \Bar{\rho}_1(n,\tau) =& \rho_1(n)\cdot(1-e^{-\tau/\tau_{SE}})\\
\end{aligned}    
\end{equation*} 
A good order of magnitude for the time interval between two consecutive trains of pulses is $\tau_{SE}$. However, after one exponential decay time, about \SI{37}{\percent} of the atoms initially in the excited state will still be excited. Being in the ``\textit{wrong}" initial state, these particles will be warmed up instead of cooled down by the second train of pulses. 

\begin{figure}
	\centering
	\includegraphics[width=0.95\linewidth]{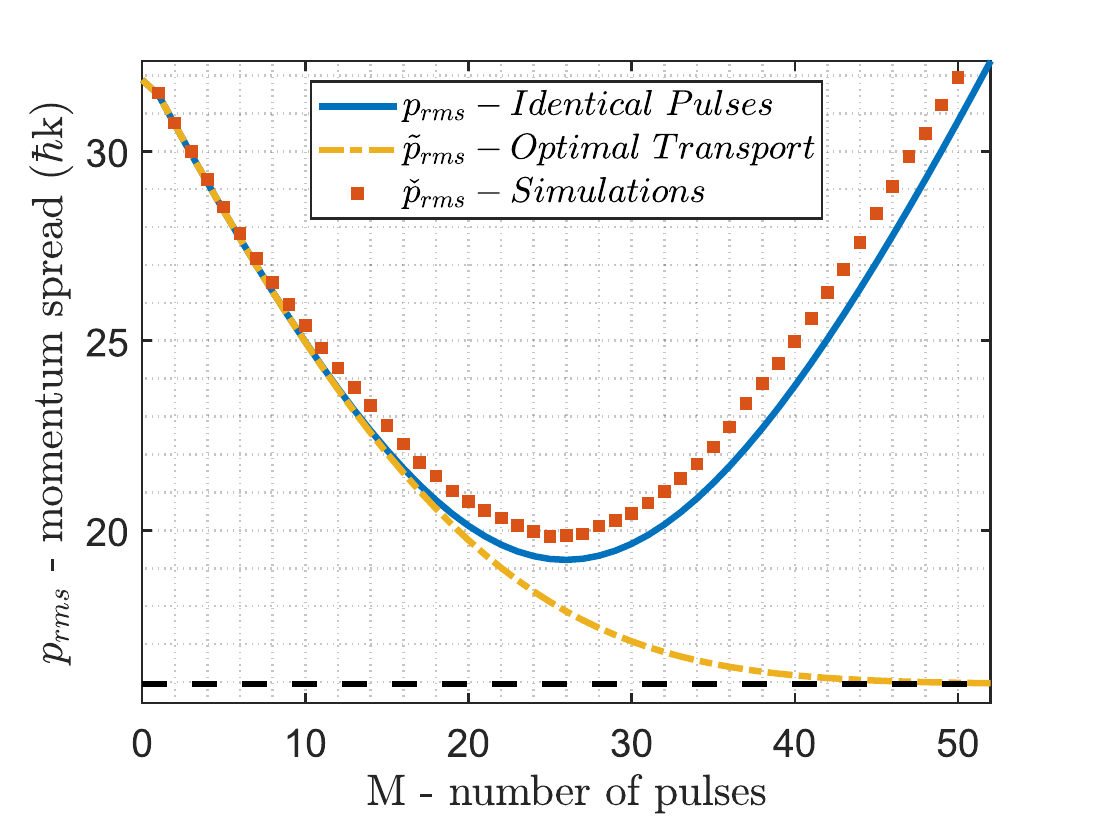}% Here is how to import EPS art
    \caption{\label{fig:sigmaVsN} Evolution of the momentum rms as function of the number of pulses as predicted analytically for the \textit{Identical Pulses} ($p_{rms}$, see eq.~(\ref{eq:analytical2}); solid blue line) and \textit{Optimal Transport} ($\tilde{p}_{rms}$, see eq.~(\ref{eq:notchedanalytical}); yellow dot--dashed line) methods and as simulated numerically for \textit{Identical Pulses} including modelling of the laser, spontaneous emission, photo--ionization and annihilation ($\Check{p}_{rms}$, see eq.~(\ref{eq:pcheck}); red squares). The initial standard deviation of the Maxwell-Boltzmann distribution is $p_{rms,0}$ = 31.88$\;\hbar k$ (corresponding to \SI{300}{\kelvin} for positronium \cite{Zimmer2021}). The black dashed horizontal line features the $p_{rms,0}/2$ level.}
\end{figure}

\begin{figure}
	\centering
\includegraphics[trim=85 260 85 250,clip,width=0.95\linewidth]{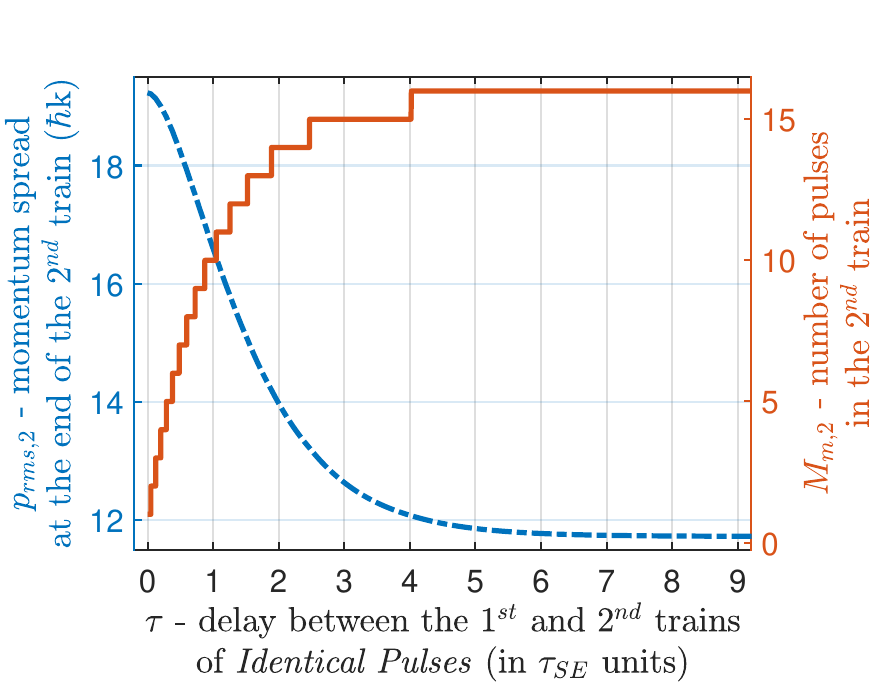}
\caption{{\label{fig:SEoptimisation}} Optimization of the delay $\tau$ between the first and second trains of \textit{Identical Pulses}. (dot-dashed blue line) Minimum standard deviation $p_{rms,2}$ (see section~\ref{sec:multipletrains}) achievable at the end of a 2$^{nd}$ train of $M_{m,2}$ pulses (solid red line) as function of the delay $\tau$ between the 1$^{st}$ and 2$^{nd}$ trains of \textit{Identical Pulses}. $p_{rms,2}(\tau=0) = p_{rms,1} = 19.23\;\hbar k$ is the minimum value reached after a first \textit{Identical Pulses} with M$_m$ of pulses (see Fig.~\ref{fig:sigmaVsN}).}
\end{figure}

To study the influence of the delay $\tau$ between two consecutive \textit{Identical Pulses} or \textit{Optimal Transport} trains of pulses, we introduce spontaneous emission and compute the momentum spread after two trains of pulses for each value of $\tau$. The momentum rms $p_{rms,2}$ at the end of the second train with $M_{m,2}$ pulses is computed with equation~(\ref{eq:hyperbola2}) applied to $\Bar{\rho}_0(n,\tau)$ and $\Bar{\rho}_1(n,\tau)$. As an example, we start from a Maxwell-Boltzmann distribution with $p_{rms,0}$ = 31.88$\;\hbar k$ with all atoms in the ground state, the first train is made of $M_{m,1} = 26$ pulses after dessymetrization yielding a minimum momentum rms $p_{rms,1} = 19.23\;\hbar k$ (see Fig.~\ref{fig:sigmaVsN}). The optimal number of pulses in the second train $M_{m,2}$ after dessymetrization depends on $\tau$. The result for the \textit{Identical Pulses} is presented in Fig.~\ref{fig:SEoptimisation}. $p_{rms,2}$ is the momentum rms reached at the end of the second train of pulses. When $\tau$ increases, there is more time for atoms in the excited state to relax to the ground state. As a result, when the second train starts, less atoms are in the ``wrong" state and $p_{rms,2}$ continuously decreases as function of $\tau$. Fig.~\ref{fig:SEoptimisation} suggests that a value of $\tau \simeq 4\;\tau_{SE}$ is a good compromise between shortening the process as much as possible and reaching the lowest possible rms at the end of the second train of pulses. For $\tau = 4\;\tau_{SE}$, the highest number of pulses in the second train is reached and the momentum rms after the second train of pulses is less than $1\;\hbar k$ away from the value that would be reached for an infinitely long delay between the two trains. Considering that with the first train of pulses, the rms dropped from 31.88 to 19.23$\;\hbar k$, if $\tau = 4\;\tau_{SE}$, 12.65$\;\hbar k$ have been removed in 4$\;\tau_{SE}$ which is a factor 6 higher than what can be achieved with traditional laser cooling. Considering a reduction of the momentum rms by approximately 7$\;\hbar k$ with the second train, if the delay between the second and the third train of pulses is still 4$\;\tau_{SE}$, a total reduction of 20$\;\hbar k$ is achieved in 8$\;\tau_{SE}$. For momentum distributions with larger initial spread, the gain would be even larger.

By comparison with traditional Doppler cooling, this result sheds light on the limit of efficiency of coherent laser cooling with trains of ultrashort laser pulses. Indeed, traditional Doppler cooling allows to reduce the momentum spread by 1$\;\hbar k$ in 2$\;\tau_{SE}$ at best. If the reduction of momentum spread induced is smaller than this rate, the proposed coherent cooling scheme is no longer competitive and should not be used. In case of two- and three-dimensional cooling, it is advisable to change the direction of cooling after this limit is reached.

% The experimental feasibility of the two schemes presented is beyond the scope of this article. However, we would like to make the following points. 
Concerning the \textit{Identical Pulses}, we note that the use of a dessymetrization step can be avoided in the case of a pulsed source of atoms for which it is reasonable to assume a spatial distribution of momentum such that the particles with highest absolute momentum are farthest from the spatial center of the ensemble. In this situation, the same momentum rms reduction is obtained by using pairs of identical pulses synchronized to reach the center of the ensemble of particles at the same time. In this situation, there is no need to manipulate classes of momentum with $\hbar k$ resolution. As for \textit{Optimal Transport}, in the particular case of positronium, the frequency shift associated with a momentum change of 1$\;\hbar k$ amounts to $\simeq$ \SI{6.15}{\giga\hertz}. As demonstrated with cutting--edge setups \cite{Metcalf:13,Ma:20}, it is possible to perform pulse shaping with the required resolution on a \SI{1}{\tera\hertz} large spectral bandwidth.

In the next section, we report on numerical simulations of the \textit{Identical Pulses} including a realistic description of the laser, spontaneous emission and photo-ionization from the excited state.

\section{Numerical simulations}

We use the Quantum Toolbox in Python to simulate the evolution of the ensemble of atoms. The behaviour of the  $\hat\rho$ density operator is described as an open quantum system using the Lindblad master equation \cite{Manzano20}:
\begin{equation}
    \dx{\hat\rho}{t} = [\hat{H},\hat{\rho}] + \sum_i \left(\hat L_i\hat\rho \hat L_i^\dag -\frac{1}{2}\left\{\hat L_i^\dag \hat L_i,\hat\rho\right\}\right)
\end{equation}
% where $[.,.]$ and $\{.,.\}$ denote the commutator and anticommutator operations, respectively. 
The unitary part of the transformation is described by Hamiltonians in the momentum space $\mathcal{H}_{V}$ and internal state space $\mathcal{H}_{I}$, which are composed into the Hamiltonian $\hat H = \hat H_{\mathrm{I}}\otimes \hat H_{\mathrm{V}}$. Additionally, the dissipative behaviour is modelled with Lindblad operators $\hat L_i$ with corresponding coupling strengths $\Gamma_i$. In $\hat H_{\mathrm{I}}$, we put the atomic levels $\ket{g}$ and $\ket{e}$. The 1-dimensional momentum space is discrete and sampled on multiple integers of $1\hbar k$. In this situation, spontaneous emission can be properly described by redistributing the decaying excited states to the ground state with three different momenta as described previously \cite{Bloch08}. This discrete momentum space allows to properly describe the dynamics of the system for rms width large in front of $\hbar k$. The inclusion of sub-Doppler effects would however require to use a higher sampling rate of this space.  %The change in wavenumber $k$ due to a frequency-sweep is neglected. 
Furthermore, the Hamiltonian is transformed to the rotating frame and is written as 
\begin{equation}\label{eq:H-cmt}
    \begin{split}
    \hat H = \hbar\sum_{n=-n_{bin}}^{n_{bin}} \Big( &n^2\omega_{rec.}\ket{g,n}\bra{g,n}\\
    &+\big(\omega_0 -\omega_L(t) + n^2\omega_{rec.} \big)\ket{e,n}\bra{e,n}\\
    &-\Omega(t)\ket{g,n}\bra{e,n\pm 1}\\
    &-\Omega(t)\ket{e,n\pm 1}\bra{g,n}\Big)
    \end{split}
\end{equation}
where $\Omega(t) = \frac{d\cdot E(t)}{\hbar}$ is the instantaneous Rabi frequency, $E(t)$ is the laser field envelope, $d$ the dipole moment of the transition and $\omega_0$ is the resonant angular frequency of the transition. The instantaneous angular laser frequency is given by $\omega_L(t)$, and $\omega_{rec.} = \frac{\hbar k^2}{2m} $ is the laser recoil frequency. Two (dead) states $\ket{ph.}$ and $\ket{ann.}$ can be added to $\hat H_{\mathrm{I}}$ to take into account losses such as photoionisation (ph.) and annihilation (ann.). These effects as well as spontaneous emission are included similarly to other work in literature\cite{Bartolotta18,Zimmer2021}, using the following Lindblad operators:

\begin{gather}
    \begin{split}
        \hat L_{SE} = \sqrt{\Gamma_{SE}}\sum_{n=-n_{bin}+1}^{n_{bin}-1}\big(\sqrt{0.6}\ket{g,n}\bra{e,n}\\
            +\sqrt{0.2}\ket{g,n-1}\bra{e,n} \\
            + \sqrt{0.2}\ket{g,n+1}\bra{e,n}\big)
    \end{split}\\
    \begin{split}
        \hat L_{ph.} = \sqrt{\Gamma_{ph.}}\sum_{n=-n_{bin}}^{n_{bin}}\ket{ph.,n}\bra{e,n}
    \end{split}\\
    \begin{split}
        \hat L_{ann.} =\sqrt{\Gamma_{ann.}}\sum_{n=-n_{bin}}^{n_{bin}}\ket{ann.,n}\bra{g,n}
    \end{split}
\end{gather}
In our simulations, $\omega_{rec.}$ was set to zero to limit computing time. This approximation has a negligible impact on the results. In the following, we present results in the case of positronium for which $\ket{g} = \ket{1S}$ and $\ket{e} = \ket{2P}$. $\Gamma_{ph.} = \sigma_{ph.}\frac{I(t)}{\hbar\omega_0}$ is the photoionisation rate and is related to the photoionisation cross-section $\sigma_{ph.} = \SI{2.6e-22}{\meter^{2}}$, $I(t)$ is the laser intensity, $\Gamma_{ann.} = \frac{1}{142\mathrm{ns}}$ is the annihilation rate in the $\ket{1S}$ state, and $\Gamma_{SE} = \frac{1}{3.2\mathrm{ns}}$ is the spontaneous emission rate.

The momentum spread $\Check{p}_{rms}$ is quantified by the standard deviation of the velocity distribution. In our studies, $\Check{p}_{rms}$ refers to the standard deviation of only the ground and excited states.
\begin{equation}\label{eq:pcheck}
    \Check{p}_{rms} = \sqrt{\tr{\ptr{\hat p\hat\rho}{I}}^2 - \tr{\ptr{\hat p \hat\rho}{I}^2}}
\end{equation}
where tr[.] denotes the trace of an operator, $\mathrm{tr}_{\mathrm{I}}$[.] denotes the partial trace of an operator over the internal (g,e) state space $\mathcal{H}_I$, and $\hat p$ is the momentum operator. 

We study the \textit{Identical Pulses}. Initially, the Doppler profile is ideally dessymetrized by transferring the population with strictly positive velocity components to the excited state as if they had absorbed 1$\;\hbar k$. Therefore, the numerical simulations do not take into account any limitations of this scheme due to the dessymetrization step. %In principle, the latter can be skipped under the reasonable assumption that the momentum distribution is spatially distributed so that the classes of momentum with highest absolute value can be excited by the laser before the classes of momentum with lowest momentum. 
To reach high population transfer efficiency with the pulses following the dessymetrization pulse, we choose to follow the Adiabatic Rapid Passage approach \cite{Melinger:92,Malinovsky:01} with chirped Gaussian laser pulses, which allows to reach close to \SI{100}{\percent} with robust pulse parameters. In this \textit{Identical Pulses} scheme, all pulses after the dessymetrization pulse are identical with a pulse envelope of the electric field $E(t)$ and instantaneous frequency $\omega_{L}(t)$:
\begin{equation}
    \begin{aligned}
        E(t) &= E_{0}e^{-2\ln{2}\frac{t^2}{\tau^2}}\\
        \omega_{L}(t) &= \omega_0 + \alpha t\\
    \end{aligned}
\end{equation}
where $E_{0}$ is the amplitude of the electric field, $\tau$ is the pulse duration defined as the full-width at half-maximum of the intensity profile and $\alpha$ is the linear chirp rate. Other pulse shapes are possible \cite{Lu:07,Guerin:11} but we choose to implement the simplest possible. The following parameters are used in the simulations: $\tau =$ \SI{7.07}{\pico\second} and $\alpha =$ 2$\pi \left(\SI{250}{\giga\hertz\per\pico\second}\right)$. The laser spectral width defined as the full-width at half-maximum of the spectral intensity is $\delta\nu = \frac{\sqrt{4(\ln{2})^2+\alpha^2\tau^4}}{2\pi\tau} \simeq \SI{1.8}{\tera\hertz}$. The peak intensity is $E_{0} =$ \SI{1.5}{\kilo\watt\per\square\centi\meter} which yields a peak Rabi frequency $\Omega_0 = 2\pi\left(\SI{150}{\giga\hertz}\right)$ with a dipole strength d = 1.97\;Debye \cite{Zimmer2021}. In Fig.~\ref{fig:sigmaVsN}, the result of the numerical simulations ($\Check{p}_{rms}$, red squares) shows good quantitative agreement with the evolution predicted by equation~\ref{eq:hyperbola2} ($p_{rms}$, solid blue curve). The minimum reached in the simulation is slightly higher than predicted by equation~\ref{eq:hyperbola2} and the number of pulses used to reach this minimum is very close to the predicted one. $\Check{p}_{rms}$ is larger than $p_{rms}$ which we attribute to incomplete population transfer. The population found in the "wrong" state at the beginning of each pulse is warmed up instead of being cooled. Since population is very efficient, in first approximation once a population is in the "wrong" state, it stays in the "wrong" state and the number of atoms in the "wrong" states increase with the number of pulses.

\begin{figure}
	\centering
	\includegraphics[width=0.95\linewidth]{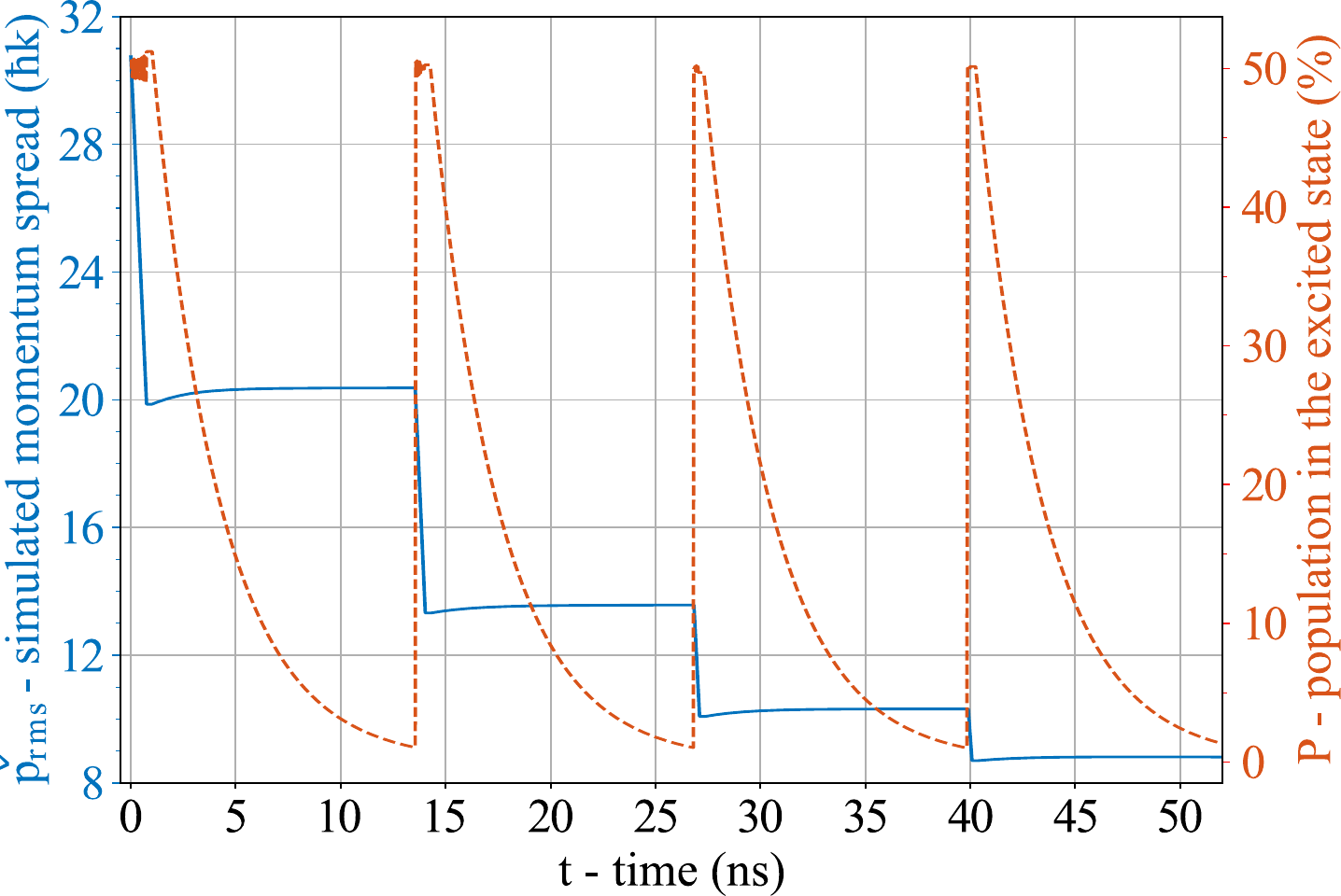}
    \caption{\label{fig:trainoftrains} $\Check{p}_{rms}$: momentum rms (solid blue line, left axis) in $\hbar k$ units simulated as function of time for a sequence of four \textit{Identical Pulses} with optimized number of pulses. The delay between two consecutive trains is  $4\;\tau_{SE}$. P: population in the excited state (dashed red line, right axis) defined as the ratio between the excited state to the sum of the ground and excited states.}
\end{figure}

To complete this study, we simulate laser cooling with four trains of pulses. Results are presented in Fig.~\ref{fig:trainoftrains}. The delay between two consecutive trains is fixed to 4$\;\tau_{SE}$. Each train starts with an ideal dessymetrization step. The number of pulses M$_{m,i}$ in the i$^{th}$ train of pulses is optimized to minimize the momentum rms $\sigma_{i}$ at the end of the $i^{th}$ train. In the present case, M$_{m,1}$ = 26, M$_{m,2}$ = 16, M$_{m,3}$ = 9 and M$_{m,4}$ = 6. At the end of the 4$^{th}$ train, the rms reaches 8.6$\;\hbar k$ representing a gain of only 1.5$\;\hbar k$ with respect to $\sigma_{m,3}$. Therefore, in this particular case, this scheme becomes less efficient than traditional laser cooling at the end of the third train of pulses. %To overcome this limit, the use of \textit{Optimal Transport} can in principle allow to reach a momentum rms of 2$\;\hbar k$ (starting from 32) with 4 trains of pulses separated by 4$\;\tau_{SE}$.

\begin{figure*}
	\centering
\includegraphics[width=\linewidth]{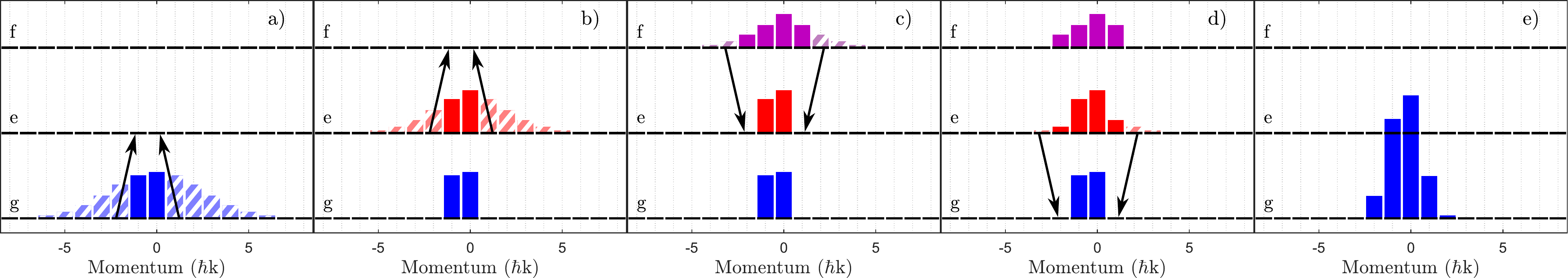}
\caption{\label{fig:3levels} Illustration of coherent laser cooling with the \textit{Optimal transport} method for a 3-level system on a distribution with an initial momentum rms of $\simeq\;$5$\;\hbar k$. Colors reflect the electronic internal level. Blue (g) corresponds to the initially populated level. Red (e) and purple (f) are other internal levels. A simple case is when (g) and (f) are two distinct levels in the ground (respectively excited) state manifold and (e) as the excited (resp. ground) level. a) Initial distribution. b) to d) intermediate distributions. e) Distribution after relaxation via spontaneous emission followed by transfer to (g). Transfer to (g) corresponds to the preparation of a single level in the ground state manifold when (g) is one of two ground state levels. When (g) is on of two excited state levels, transfer to (g) corresponds to excitation from the ground state level to (e).}
\end{figure*}

Finally, we propose that the \textit{Optimal Transport} coherent laser cooling scheme can be extended to N-level systems considering levels that are part of two manifolds only and a system initially prepared in a single state of one of the two manifolds. Fig.~\ref{fig:3levels} illustrates how \textit{Optimal Transport} can be extended to the 3-level case. With this method, eq.~\ref{eq:notchedanalytical} can be generalized to the N-level case as described for $M \geq 1$ by eq.~\ref{eq:optimaltransportNlevels} given in Appendix. For a Maxwell--Boltzmann distribution with an initial momentum rms $\tilde{p}_{rms,0}$ large in front of $N\hbar k$ and no particle initially in any of the other states, the population in the class of momentum with n$\hbar k$ when all N levels are used writes:
\begin{equation*}
    \sum_{z = 0}^{N-1}\rho_{0}(Nn+z) \simeq \frac{N}{\tilde{p}_{rms,0}\sqrt{2\pi}}\cdot\text{e}^{-\frac{N^2(n\hbar k)^2}{2\tilde{p}_{rms,0}^2}}
\end{equation*}
which corresponds to a Maxwell--Boltzmann distribution with a momentum rms divided by a factor N compared to the initial one. The precise efficiency of such multi--level coherent laser cooling goes beyond the scope of the present study. However, we note that the complexity of the required pulse shaping and the time needed to prepare the initial distribution with all particles in a single level of a manifold will most likely limit the interest of such an extension.

\section{Conclusion}
In conclusion, our study introduces a new laser cooling scheme based on trains of ultrashort pulses extending coherent laser cooling to broad line width transitions. This technique is particularly relevant for distributions with large initial momentum spread and shows that it is possible to remove several $\hbar k$ per spontaneous emission lifetime. It is particularly important to optimize the time it takes to compress the momentum spread compression in particular in view of forming a Bose Einstein Condensate \cite{Platzman94,Giatzoglou:CsIsotopes} to generate coherent $\gamma$-ray emission by coherence transfer from positronium \cite{Avetissian14} or radio-active elements \cite{Marmugi2018}. We derived analytical formulae predicting the evolution of the momentum distribution rms as function of the number of pulses in the train for two different schemes for population manipulation. The \textit{Identical Pulses} makes use of an optimal number of identical pulses. \textit{Optimal Transport} allows to realize the maximum of momentum rms reduction without entropy dissipation (a factor 2 for a 2--level system). We find that in the case studied here, a delay of 4$\;\tau_{SE}$ between two consecutive trains of pulses allows to optimize the cooling process. These results are confirmed by realistic numerical simulations. For systems with more than 2 internal levels, it is in principle possible to further increase the efficiency of this laser cooling scheme, which could be particularly interesting for molecule laser cooling. 

\section*{Acknowledgments}
We are indebted to Dr. Benjamin Bonnefont for discussions on the Monge-Kantorovich problem and Optimal Transport. This work was supported by 
% Jan Malamant
the CERN technical student program;
% Antoine Camper
the Research Council of Norway under Grant Agreement No. 303337 (PsCool), 303347 (UNLOCK), 326503 (MIR) and NorCC;
% Natali Gusakova
CERN and NTNU doctoral programs.
% Irina T Sorokina
% the Research Council of Norway under Grant Agreement No. 303347 (UNLOCK)  and No. 326503 (MIR).
\section*{Appendix}
\sectionmark

\begin{equation}
\begin{aligned}
    p_{rms,0} =& \sqrt{\sum_{n}n^2(\rho_{0}(n)+\rho_{1}(n))} \\
\end{aligned}
\end{equation}
\begin{equation}
\begin{aligned}
    M_{mc} =& \sum_{n > 0}n\left(\rho_{0}(n)+\rho_{0}(-n)\right)+\sum_{n\leq 0}\rho_{0}(n)\\
    -& \sum_{n > 0}n\left(\rho_{1}(n)+\rho_{1}(-n)\right)+\sum_{n\leq 0}\rho_{1}(n)\\
\end{aligned}
\end{equation}
\begin{equation}
\begin{aligned}
    M_{ms} =& 2\sum_{n> 0}n\rho_{0}(-n)+\sum_{n\leq 0}\rho_{0}(n)\\
    -& 2\sum_{n> 0}n\rho_{1}(-n)+\sum_{n\leq 0}\rho_{1}(n)\\
\end{aligned}
\end{equation}

\begin{equation}\label{eq:optimaltransportNlevels}
\begin{aligned}
    &\tilde{p}^{2}_{rms}(M) = \sum_{n = 0}^{\floor{\frac{M}{N-1}}-1}n^2\sum_{z = 0}^{N-1}\rho_{0}(Nn+z)\\
     &+ \floor{\frac{M}{N-1}}^2\sum_{z = 0}^{M-\floor{\frac{M}{N-1}}(N-1)}\rho_{0}\left(N\floor{\frac{M}{N-1}}+z\right)\\
     &+ \sum_{n = -\floor{\frac{M}{N-1}}}^{-1}n^2\sum_{z = 0}^{N-1}\rho_{0}(Nn-z)\\
     &+ \left(\floor{\frac{M}{N-1}}+1\right)^2\sum_{z = 0}^{M-\floor{\frac{M}{N-1}}(N-1)}\rho_{0}\left(-N\left(\floor{\frac{M}{N-1}}+1\right)-z\right)\\
     &+ \sum_{n \geq \floor{\frac{M}{N-1}}+1}n^2\rho_{0}(n+M)\\
     &+ \sum_{n \leq M+1}n^2\rho_{1}(n-M)\\
     &+ \sum_{n \leq -\floor{\frac{M}{N-1}}-2}n^2\rho_{0}(n-M+1)\\
     &+ \sum_{n \geq -M}n^2+\rho_{1}(n+M-1)
\end{aligned}
\end{equation}

\nocite{*}

\bibliography{biblio}% Produces the bibliography via BibTeX.

\end{document}